\documentclass[%
  journal = jpclcd,%
  layout = twocolumn,%
  ]{achemso}
\usepackage{amsmath, amssymb}
\usepackage{mciteplus}
\newcommand{\onlinecite}[1]{\hspace{-1 ex} \nocite{#1}\citenum{#1}} 
\newcommand{\citebrackets}[1]{\mbox{(\hspace{-0.8ex} \citenum{#1})}}

\usepackage{graphicx}
\usepackage[ colorlinks,
    linkcolor={black},
    citecolor={black},
    urlcolor={black}]{hyperref}
\usepackage{braket}
\usepackage{enumitem}

\usepackage[varg]{txfonts}

\newcommand{\bigS}[1]{S_{\hspace{-0.18em}#1}}
\newcommand\eqt{\hspace{0.17em}{=}\hspace{0.17em}}

\newcommand\pt{\hspace{0.17em}{+}\hspace{0.17em}}

\newcommand\cdott{\hspace{0.13em}{\cdot}\hspace{0.13em}}
\newcommand\sd{\hspace{0.05em}}

\newcommand\deft{\hspace{0.17em}{=:}\hspace{0.17em}}

\newcommand\otimest{\hspace{0.12em}{\otimes}\hspace{0.12em}}

\newcommand{\red}[1]{\textcolor{black}{#1}}
\newcommand{\blue}[1]{\textcolor{black}{#1}}

\title{Towards \textit{GW} Calculations on Thousands of Atoms}

\author{Jan Wilhelm}\email{jan.wilhelm@basf.com}
\affiliation{Department of Chemistry, University of  Zurich, 
Winterthurerstrasse 190, CH-8057 Zurich, Switzerland}  
\alsoaffiliation{Present address: BASF SE, Carl-Bosch-Stra\ss e 38, D-67056 Ludwigshafen am Rhein, Germany}
\author{Dorothea Golze}
\affiliation{COMP/Department of Applied Physics, Aalto University, P.O. Box 11100,  FI-00076 Aalto, Finland}
\author{Leopold Talirz} 
\affiliation{Theory  and  Simulation  of  Materials, \'Ecole  Polytechnique  F\'ed\'erale  de  Lausanne,  Station 9, CH-1015  Lausanne,  Switzerland}   
\alsoaffiliation{Laboratory of Molecular Simulation, \'Ecole Polytechnique F\'ed\'erale de Lausanne, Rue de l'Industrie 17, CH-1951 Sion, Switzerland}
\author{J\"urg Hutter}\affiliation{Department of Chemistry, University of  Zurich, Winterthurerstrasse 190, CH-8057 Zurich, Switzerland}
\author{Carlo A.~Pignedoli}
\email{carlo.pignedoli@empa.ch}
\affiliation{Swiss Federal Laboratories for Materials Science and Technology (Empa), \"Uberlandstrasse 129, CH-8600 D\"ubendorf, Switzerland}  
\date{\today}

\begin{tocentry}
\includegraphics[width=1.00\textwidth]{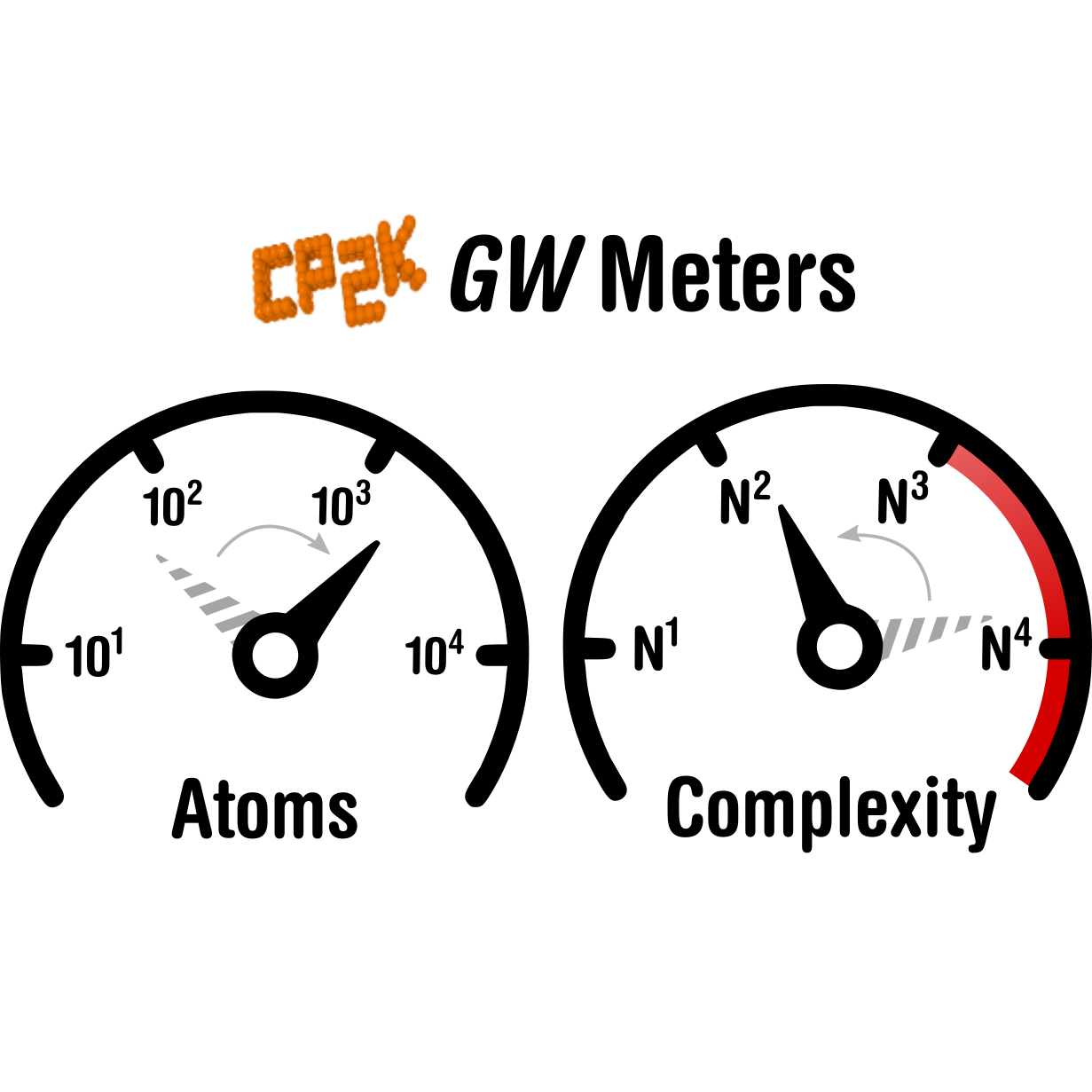}
\end{tocentry}

\begin{document}

\begin{abstract}
%
The $GW$ approximation of many-body perturbation theory is an accurate method for computing electron addition and removal energies of molecules and solids.
In a canonical implementation, however, its computational cost is $\mathcal{O}(N^4)$ in the system size $N$, which prohibits its application to many systems of interest.
We present a full-frequency $GW$ algorithm in a Gaussian-type basis,
whose computational cost scales with $N^2$ to $N^3$.
The implementation is optimized for massively parallel execution on state-of-the-art supercomputers
and is suitable for nanostructures and molecules in the gas, liquid or condensed phase,
using either pseudopotentials or all electrons.
We validate the accuracy of the algorithm on the $GW$100 molecular test set, finding
mean absolute deviations of 35\,meV for ionization potentials and 27\,meV for electron affinities.
Furthermore, we study the length-dependence of quasiparticle energies in armchair 
graphene nanoribbons of up to 1734 atoms in size, and compute the local density of states across a nanoscale heterojunction.
\end{abstract}

Electronic excitations in nanostructures
and at complex interfaces play a decisive role in several key materials challenges, such as energy conversion~\citebrackets{C3CS00007A} and digital electronics~\citebrackets{schwierz2013graphene}.
The $GW$ approximation of many-body perturbation theory~\citebrackets{Hedin,RevModPhys.74.601} is a method devised for computing the energies of charged excitations, 
which involve the addition or removal of electrons.
It accounts for the non-local, frequency-dependent screening of the interaction
between electrons, which is particularly essential where materials vary over electronic length scales.
The $GW$ spectra can be compared to photoemission spectroscopy and scanning-tunneling spectroscopy,
and form the basis for the accurate prediction of optical spectra via the Bethe-Salpeter equation ~\citebrackets{JPCLJacquemin2017}.
The good performance of the $GW$ approximation in predicting band structures of solids and, more recently,
ionization potentials and electron affinities of molecules~\citebrackets{marom2017accurate} has led to increasing interest from the chemistry community.
However, the computational complexity of the canonical $GW$ algorithm~\citebrackets{aimsgw,Blasealt,GWCP2K,BerkeleyGW} is $\mathcal{O}(N^4)$ in the system size $N$, with a substantial prefactor.
This would prohibit the study of many systems of interest, such as solid-liquid interfaces~\citebrackets{Galli}, 
large metal complexes in solution~\citebrackets{Cubanes}, metal-organic frameworks~\citebrackets{MOFs}, defect states~\citebrackets{PhysRevLett.113.136602} or $p$-$n$ junctions~\citebrackets{Bdopingribbon,ribbonheterojunctions} that require calculations on
hundreds to thousands of atoms.

In recent years, substantial efforts have therefore been devoted to reducing the computational cost of $GW$ calculations.
The prefactor has been tackled by avoiding the sum over empty states in the polarizability~\citebrackets{Umari2,Galli,bruneval2016optimized} 
as well as with a low-rank approximation of the dielectric matrix~\citebrackets{Galli,PhysRevB.81.115105}.
The size of the matrices involved can also be reduced by switching from the traditional plane-wave basis to smaller, localized basis sets~\citebrackets{Blasealt,aimsgw,GWCP2K,FEGW,molgwimpl}, which are particularly suited for molecular systems~\citebrackets{BrunevalStart,benchmarkXB,knight2016accurate,RangelAcenes}.
Others have tackled the exponent:
Foerster~\textit{et al.}~\citebrackets{Scaling} devised a cubic-scaling $GW$ algorithm in a Gaussian basis 
that exploits the locality of electronic interactions. The method has been 
applied to molecules with tens of atoms.
Liu ~\textit{et al.}~\citebrackets{liu2016cubic} implemented a variant of the cubic-scaling $GW$ space-time method~\citebrackets{AC1}, using a plane-wave basis, real-space grids and 
sophisticated minimax quadratures~\citebrackets{liu2016cubic,RPAKresse} in imaginary time and frequency.
Its linear scaling with the number of $k$-points is particularly promising for applications to large and numerically challenging periodic systems. 
Finally, Neuhauser~\textit{et al.}~\citebrackets{ONGW} 
reported a stochastic $GW$ algorithm which nominally enables
linear scaling with system size and is straightforward to parallelize.
The algorithm has been applied to a silicon nano\-cluster with one thousand atoms,
but further exploration is needed to verify that stochastic $GW$ is a useful tool for more complex systems~\citebrackets{vlcek2016stochastic}.

In this work, we present an efficient low-scaling $GW$ algorithm in a Gaussian basis
that has been optimized for massively parallel execution on state-of-the-art
supercomputers.
In comparison to plane waves, the smaller size of the Gaussian basis
together with the exploitation of sparsity in two- and three-index tensor operations
increase performance while maintaining accuracy, as we demonstrate on the
$GW$100 test set~\citebrackets{GW100}.
The algorithm is suited for nanostructures and molecules in the gas, liquid or condensed phase 
and is implemented in version 5.0 of the open-source CP2K package~\citebrackets{reviewJuerg}.

\begin{figure}[]
\centering
\includegraphics[width=0.47\textwidth]{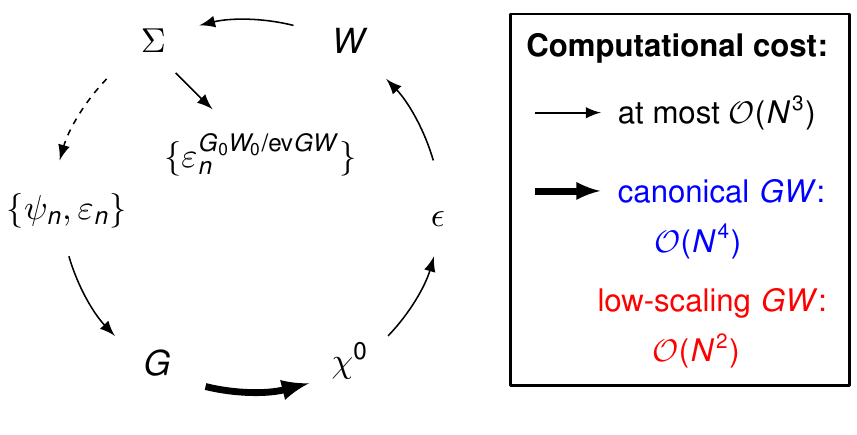}
\caption{
Sketch of $GW$ workflow. 
Regular arrows indicate operations of at most $\mathcal{O}(N^3)$ computational complexity.
The thick arrow is $\mathcal{O}(N^4)$ in canonical $GW$ (blue) and $\mathcal{O}(N^2)$ in low-scaling $GW$ (red).
The dashed arrow corresponds to eigenvalue-selfconsistent $GW$ (ev$GW$).
A detailed flowchart of the canonical and the low-scaling $GW$ algorithm can be found in the Supporting Information on page~S2.
}
\label{fig:workflow2}
\end{figure}
As sketched in Fig.~\ref{fig:workflow2}, 
the $GW$ calculation starts from a set $\{\psi_n,\varepsilon_n\}$ of single-particle orbitals~$\psi_n$ and corresponding eigenenergies~$\varepsilon_n$.
Usually, these stem from a previous Kohn-Sham density functional theory (DFT) calculation, but other starting points, 
such as Hartree-Fock and hybrid functionals, are also available.

The orbitals are expanded in the primary Gaussian-type orbitals (GTOs) $\{\phi_\mu\}$
\begin{align}
\psi_n(\mathbf{r}) = \sum_\mu C_{n\mu }\phi_\mu(\mathbf{r})\label{expGauss}
\end{align}
using the molecular orbital (MO) coefficients~$C_{n\mu }$.

Following the $GW$ space-time method, 
we proceed to computing the time-ordered single-particle Green's function 
$G(i\tau)$
in imaginary time:
\begin{align}
\begin{split}
G_{\mu\nu}(i\tau)= \left\{ 
\begin{array}{ll}
  i\sum\limits_n^\text{occ} C_{n\mu}C_{n\nu}\exp(\varepsilon_n\tau)\,, &  \tau >0\,,
\\[0.5em]
 - i\sum\limits_n^\text{virt} C_{n\mu}C_{n\nu}\exp(\varepsilon_n\tau)\,, &  \tau <0\,.
\end{array}
\right.
\end{split}
\label{Greensf}
\end{align}
A key step in the algorithm is computing the irreducible polarizability~$\chi^0(i\tau)\eqt -iG(i\tau)G(-i\tau)$.~\citebrackets{AC1}
Building on previous work~\citebrackets{cubicRPAcp2k},
$\chi^0$ is obtained in an auxiliary Gaussian basis $\{\varphi_P\}$~\citebrackets{RIWeigend,MauroRI,EMSL} that is 
designed to span the product space $\{\psi_i\}\otimest\{\psi_a\}$ of occupied and unoccupied orbitals,
and is typically two to three times larger than the corresponding primary basis $\{\phi_\mu\}$.
The matrix~${\chi}^0_{PQ}(i\tau)\eqt \braket{\varphi_P|\chi^0(i\tau)|\varphi_Q}$ is calculated as
\begin{align}
\begin{split}
{\chi}^0_{PQ}(i\tau)=-i&\sum\limits_{\mu\sigma}\sum\limits_\lambda (\lambda\sigma P) G_{\mu\lambda}(i\tau)
\\[0.5em]
&\times\sum\limits_\nu (\mu\nu Q) G_{\nu\sigma}(-i\tau)\,,\label{NsquareRPA}
\end{split}
\end{align}
where the three-center overlap tensors 
\begin{align}
(\nu\mu P) = {\int} d\mathbf{r}\; \phi_\nu(\mathbf{r}) \phi_\mu(\mathbf{r})\varphi_P(\mathbf{r})\label{3coverl}
\end{align}
are computed analytically~\citebrackets{cubicRPAcp2k}.

Since the overlap tensors~$(\mu\nu P)$ vanish unless the GTOs~$\phi_\mu$, $\phi_\nu$ and~$\varphi_P$ are centered on nearby atoms,
their size grows only linearly with the system size~$N$. The computational cost of Eq.~\eqref{NsquareRPA} is therefore $\mathcal{O}(N^2)$ \red{without the requirement of sparse density matrices or additional localization techniques.}
The overlap tensor in Eq.~\eqref{3coverl} can be understood as deriving from the resolution of the identity (RI) with the \emph{overlap metric} \red{(RI-SVS)}~\citebrackets{RI3,Schurkus,Blaseoverlapmetric}.
We note that the popular RI with the \emph{Coulomb metric} \red{(RI-V)}~\citebrackets{RI3} \red{converges faster with the size of the RI basis, but} does not lead to sparsity in \eqref{3coverl} and would thus provide no advantage over the canonical implementation (see supporting information).

Although the cost of the matrix-matrix multiplication in Eq.\,\eqref{Greensf} and all following matrix operations in Eqs.~\eqref{chipseudoimtime}, \eqref{dielfunc}, \eqref{Coulomb}, \eqref{Wc} scale cubically with system size, Eq.~\eqref{NsquareRPA} remains the computational bottleneck even for the largest systems addressed in this work. 
Therefore, the computation of the polarizability from Eq.~\eqref{NsquareRPA} has been optimized for massive parallelism~\citebrackets{cubicRPAcp2k} using the DBCSR library for sparse matrix-matrix multiplications~\citebrackets{dbcsr}.
We proceed by including the non-orthogonality of $\{\varphi_P\}$ in $\boldsymbol{\tilde{\chi}}^0(i\tau)$,
\begin{align}
\boldsymbol{\tilde{\chi}}^0(i\tau)=\mathbf{S}^{-1}\boldsymbol{\chi}^0(i\tau)\mathbf{S}^{-1}\label{chipseudoimtime}
\end{align}
via the overlap matrix
\begin{align}
\bigS{PQ}\eqt {\int} d\mathbf{r}\, \varphi_P(\mathbf{r})\varphi_Q(\mathbf{r})\,.
\end{align}

Following the route of the $GW$ space-time method~\citebrackets{AC1}, 
the polarizability~$\boldsymbol{\chi}^0(i\tau)$ is transformed to imaginary frequencies via a cosine transform on the minimax grid,~\citebrackets{liu2016cubic} and the symmetric dielectric function~$\epsilon(i\omega)$ 
is computed by~\citebrackets{periodicGWCP2K}
\begin{align}
\boldsymbol{\epsilon}(i\omega) =
 \mathbf{1}- \mathbf{L}^\text{T}\boldsymbol{\tilde{\chi}}^0(i\omega)\mathbf{L}  
\label{dielfunc}
\end{align}
where $\mathbf{L}$ denotes the Cholesky decomposition of the Coulomb matrix~$\mathbf{V}$,
\begin{align}
\mathbf{V}= \mathbf{L}\mathbf{L}^\text{T}\,,\hspace{1em} 
V_{PQ}={\int} d\mathbf{r}\sd d\mathbf{r}'\, \varphi_P(\mathbf{r}) \, \frac{1}{|\mathbf{r}-\mathbf{r'}|}\, \varphi_Q(\mathbf{r}') \,.\label{Coulomb}
\end{align} 
For molecules, the Coulomb matrix is computed analytically~\citebrackets{DorotheaIntegrale} and for periodic systems numerically by Ewald summation~\citebrackets{Ewald}, as commonly used in wavefunction correlation methods~\citebrackets{MauroMP2Grad,vladimir,MauroMP2RPAlargescale}.
We note that the algorithm supports both aperiodic and periodic simulation cells in the $\Gamma$-only approach~\citebrackets{periodicGWCP2K}. 
For periodicity in three dimensions, a correction scheme is available to accelerate the convergence with supercell size~\citebrackets{periodicGWCP2K}.

The screened interaction $W(i\omega)=\epsilon^{-1}(i
\omega)V= V \pt W^c(i\omega)$ is split into the bare Coulomb interaction
and the correlation contribution, and the latter 
is obtained as~\citebrackets{periodicGWCP2K}
\begin{align}
\mathbf{W}^\text{c}(i\omega) = \mathbf{L}\left[\boldsymbol\epsilon^{-1}(i\omega) -\mathbf{1}\right]\mathbf{L}^\text{T}\,,\label{Wc}
\end{align}
where the symmetric, positive definite $\boldsymbol{\epsilon}(i\omega)$ is inverted efficiently by Cholesky decomposition.
A cosine transform brings $W^c(i\omega)$ back to imaginary time. 

This completes the ingredients for the $GW$ self-energy $\Sigma(i\tau)\eqt iG(i\tau)W(i\tau) \deft\Sigma^\text{x}\pt\Sigma^\text{c}(i\tau)$. 
In the following, we restrict the treatment to $GW$ schemes without orbital updates, such as $G_0W_0$ and eigenvalue self-consistent $GW$ (ev$GW$).
Computing the quasiparticle energies for~$N_{GW}$ orbitals~$\psi_n$
then only requires the $N_{GW}$ corresponding 
diagonal matrix elements
$\Sigma_n(i\tau) \eqt \braket{\psi_n|\Sigma(i\tau)|\psi_n}$.
For reasons of computational efficiency, 
we compute the diagonal elements directly, yielding the correlation self-energy
\begin{align}
\Sigma_n^\text{c}(i\tau) =i\sum_{\nu P}\sum_{\mu}G_{\mu\nu}(i\tau) (n\mu P)\sum_Q\tilde{W}^\text{c}_{PQ}(i\tau)(Q\nu n)\,,
\label{corrSEtaurewrite}
\end{align}
where 
$
\tilde{\mathbf{W}}^\text{c}(i\tau)\eqt\mathbf{S}^{-1}\mathbf{W}^\text{c}(i\tau)\mathbf{S}^{-1}
$,
and the static exchange self-energy
\begin{align}
\Sigma^\text{x}_n =  -
\sum_{\nu P} 
\sum_{\mu} D_{\mu\nu}(n\mu P) \sum_Q \tilde{V}_{PQ} (Q\nu n)\label{Sx}\,,
\end{align}
where~$D_{\mu\nu}\eqt\sum_n^\text{occ} C_{n\mu}C_{n\nu}$ and $\tilde{\mathbf{V}}=\mathbf{S}^{-1}\mathbf{V}\mathbf{S}^{-1}$.
The computational complexity of Eq.~\eqref{corrSEtaurewrite} and~\eqref{Sx} is $\mathcal{O}(N_{GW}N^2)$, since $(n\mu P)\eqt {\sum_\nu} C_{n\nu} (\nu\mu P)$ vanishes if $\phi_\mu$ and $\varphi_P$ are centered on atoms far apart from each other, introducing sparsity. 
In order to compute quasiparticle energies,
$\Sigma^\text{c}_n(i\tau)$ is transformed to imaginary frequencies by a sine and cosine transform.~\citebrackets{liu2016cubic} It is then evaluated on the real frequency axis $\Sigma^\text{c}_n(\varepsilon)$ by analytic continuation using a Pad\'{e} interpolant of $\Sigma^\text{c}_n(i\omega)$~\citebrackets{GW100,liu2016cubic}.
The $G_0W_0$ quasiparticle energies~$\varepsilon_n^{G_0W_0} $ are obtained by replacing the DFT exchange-correlation contribution~$v^\text{xc}_n$ with the self-energy,
\begin{align}
\varepsilon_n^{G_0W_0} = \varepsilon_n -v^\text{xc}_n 
+\Sigma^\text{x}_n+\text{Re}\,\Sigma^\text{c}_n(\varepsilon_n^{G_0W_0})
\label{qpeq}
\end{align}
and solving Eq.~\eqref{qpeq} iteratively for~$\varepsilon_n^{G_0W_0}$ via Newton-Raphson.
For eigenvalue-selfconsistent $GW$, the quasiparticle energies then replace the DFT levels in  Eqs.\,\eqref{Greensf} and the $GW$ cycle of Fig.~\ref{fig:workflow2} is repeated until self-consistency in the quasiparticle energies~$\varepsilon_n^\text{ev\textit{GW}}$ is achieved. 
\begin{figure}[t]
\centering
\includegraphics[width=0.47\textwidth]{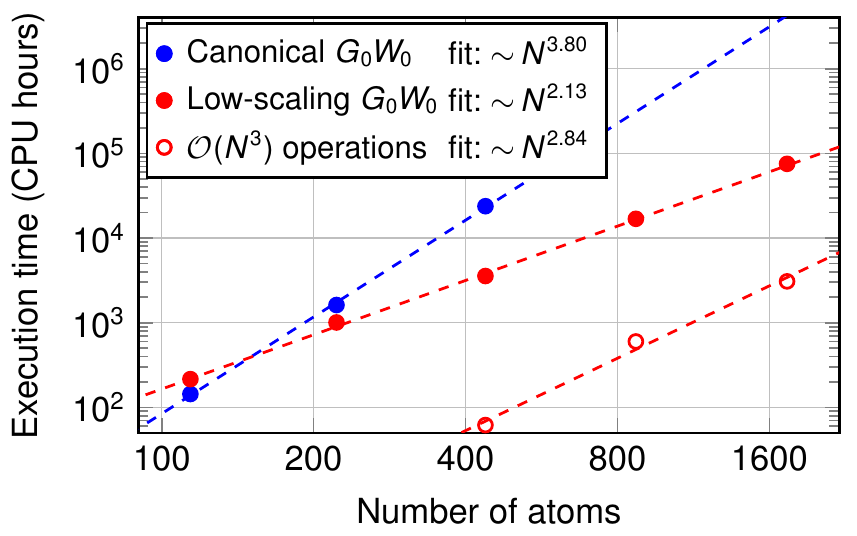}
\caption{
Scaling of $G_0W_0$ execution time with number of atoms.
The canonical algorithm~\citebrackets{GWCP2K} (blue dots) 
is compared against the low-scaling algorithm (red dots) and its fraction of cubically-scaling operations (red circles).
Dashed lines are two-parameters least-squares fits of prefactor and exponent.
The systems under study are graphene nanoribbons as shown in Fig.~\ref{computed_zz_AC_gap}.
For the largest system containing 1734 atoms, we compute
2883 occupied molecular orbitals and 30195 virtual ones, using 80940 auxiliary basis functions. This calculation was performed on 14400 CPUs on a CRAY XC40 machine.
}
\label{scaling}
\end{figure}
Fig.~\ref{scaling} illustrates how the computational cost of the algorithm
scales with the number of atoms~$N$ 
for a technologically relevant test system of graphene nanoribbons,
which is discussed in more detail below.
The total execution time of the canonical $G_0W_0$@PBE implementation~\citebrackets{GWCP2K} (blue) scales with~$N^{3.8}$, and constraints in computation time and memory prohibit us from going beyond 500 atoms.
\red{The low-scaling algorithm (red) becomes superior between 100 and 200 atoms, }
is already a factor of 8 faster at 438 atoms than the canonical implementation, and allows to reach much larger system sizes on the same computer architecture  (1734 atoms and 5766 electrons in this example).

We stress that the cost of the low-scaling algorithm scales like~$N^{2.1}$ with the number of atoms~$N$ in the range considered here, 
since the cubic-scaling steps (red circles) have a much smaller prefactor than the $\mathcal{O}(N^2)$ evaluation of  Eq.~\eqref{NsquareRPA} \red{involving sparse tensor operations}.
\red{In this regime,} we expect the algorithm to be particularly efficient for low-dimensional systems, such as 2d materials or 1d polymers and wires, as well as for systems with a local electronic structure, such as molecules in solution, 
which give rise to sparse density matrices and Green's functions~\citebrackets{BaerHG}. 
\red{For very large systems, the cubic-scaling steps will dominate and sparsity
becomes irrelevant.} By extrapolating the data shown in Fig.~\ref{scaling}, we estimate the cross-over from quadratically-dominated to cubically-dominated to occur at $\approx$\,$ 3\cdott 10^4$ atoms ($\approx$\,$10^5$ electrons) for the systems under study.
\red{For very small systems, all three-center integrals have to be retained  
and the larger RI-SVS basis puts the low-scaling $GW$ at a slight disadvantage compared to the canonical algorithm using RI-V.}

The accuracy of the low-scaling $GW$ algorithm is validated
on the $GW$100 set by van Setten~\textit{et al.}~\citebrackets{GW100}
We compute the energies of the highest occupied molecular orbital (HOMO), or \emph{ionization potential},
and the lowest unoccupied molecular orbital (LUMO), or \emph{electron affinity},
at the $G_0W_0$@PBE level for all molecules in the set.
All values are reported in the Supporting Information on pages~S3/S4
and compared to reference values
from FHI-aims~\citebrackets{GW100,aimsgw}, an all-electron code using
numerical, atom-centered basis functions.
We find that HOMO energies match within 30\,meV for 74 out of 100 molecules, while LUMO energies match within 30\,meV for 87 molecules.

For comparison, we note that HOMO energies from FHI-aims and VASP~\citebrackets{liu2016cubic}, a plane-wave code implementing the projector augmented wave method~\citebrackets{PAW}, have a mean absolute deviation (MAD) of 60\,meV on a subset of $GW$100~\citebrackets{maggio2016gw}, while we find a MAD of 35\, meV between FHI-aims and our algorithm (on $GW$100 excluding BN, O$_3$, BeO, MgO, CuCN and Ne).
We conclude that our implementation is suitably accurate and continue by discussing its application to large systems.

\begin{figure}
\centering
\includegraphics[width=0.47\textwidth]{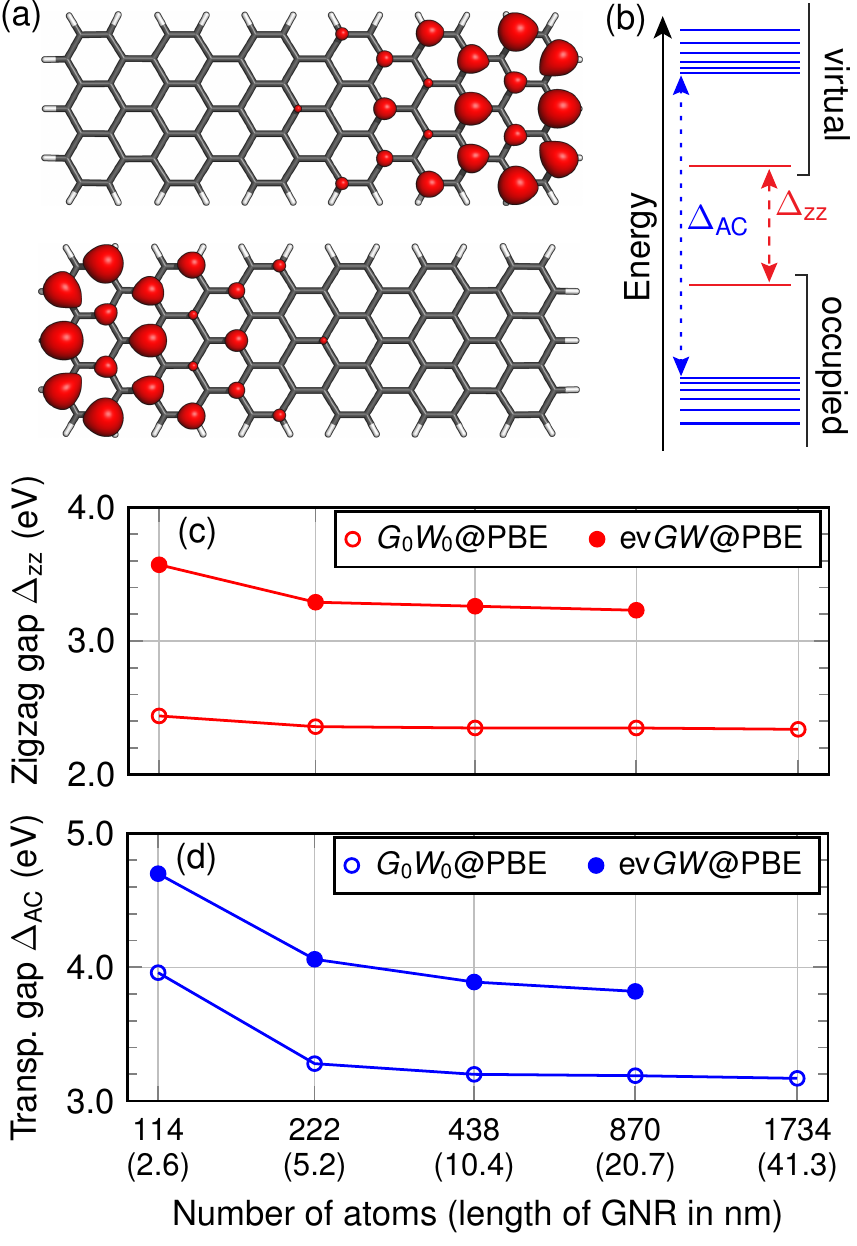}
\caption{(a) Molecular geometry of 6-anthene together with the zigzag edge states obtained from DFT. \red{All edge carbon atoms are passivated by a single hydrogen atom}. Shown in red are isosurfaces of constant probability density for the HOMO (top) and LUMO (bottom) of the spin-up channel.
(b) Sketch illustrating the corresponding spectrum with the HOMO$-$LUMO gap~$\Delta_\text{zz}$ between zigzag (zz) states 
and the transport gap~$\Delta_\text{AC}$  (\mbox{HOMO-1$-$}LUMO+1) between bulk states.
(c) Zigzag gap~$\Delta_\text{zz}$ and (d) transport gap~$\Delta_\text{AC}$  of anthenes with a horizontal length from 2.6\,nm to 41.3\,nm. 
}
\label{computed_zz_AC_gap}
\end{figure}

We start by studying anthenes, graphene nanoribbons (GNRs) of seven carbon atoms width, as depicted in Fig.~\ref{computed_zz_AC_gap}\,(a).
Recent advances in on-surface chemistry have enabled the bottom-up fabrication of these 
GNRs with atomic precision~\citebrackets{Cai2010}, and their electronic structure
has been investigated in detail by scanning tunneling spectroscopy~\citebrackets{RuffieuxAGNR7,wang2016giant}.
For these particular GNRs, HOMO and LUMO are found to be localized at the zigzag edges of the ribbons,
as depicted in Fig.~\ref{computed_zz_AC_gap}\,(a),
while the remaining frontier orbitals delocalize along the ribbon.
One therefore distinguishes the zigzag gap~$\Delta_\text{zz}$ between edge-localized HOMO and LUMO states, 
and the armchair gap~$\Delta_\text{AC}$ between the delocalized HOMO-1 and LUMO+1,
as sketched in Fig.~\ref{computed_zz_AC_gap}\,(b). 
Since only the delocalized states are available for charge transport
along the ribbon, 
$\Delta_\text{AC}$ is also termed the 
transport gap.
We compute~$\Delta_\text{zz}$ and~$\Delta_\text{AC}$ for anthenes containing up to 1734 atoms, see Fig.~\ref{computed_zz_AC_gap}\,(c) and (d).

As expected from their highly localized nature, and in agreement with previous work~\citebrackets{wang2016giant}, the zigzag gap converges 
quickly with length.
First, we note that the converged $G_0W_0$@PBE value of $\Delta_\text{zz} \eqt 2.4\,$eV is significantly lower than the 2.8\,eV reported in Ref.~\citenum{wang2016giant}, where the frequency-dependence of the polarizability was approximated by a plasmon-pole model. 
This is in line with findings for molecules 
in $GW$100~\citebrackets{GW100} and indicates that plasmon-pole models should be 
avoided in future studies of localized states in GNRs.
Secondly, self-consistency in the eigenvalues leads to a substantial increase of the gap to 3.2\,eV.
This observation \red{is easily understood} by considering that the tiny PBE Kohn-Sham gap of 0.6\,eV 
gives rise to a strong screening of this localized state in the interaction $W_0$ that is suppressed by the larger $GW$ gap in subsequent self-consistency iterations.
\red{Techniques for improving the DFT starting point include the use of hybrid density functionals with adequate fractions of Hartree-Fock exchange ~\citebrackets{aimsazabenzenes,PhysRevB.86.041110}.}

The low-scaling algorithm also allows us to study the convergence of the transport gap with GNR length, 
which requires significantly longer GNRs due to the delocalized nature of the involved electronic states.
As shown in Fig.~\ref{computed_zz_AC_gap}\,(d),
the transport gap saturates at a value of $\Delta_\text{AC}\eqt3.2\,$eV ($G_0W_0@$PBE). 
Again, this value is significantly smaller than the value of 3.8\,eV reported in early $G_0W_0$@LDA calculations~\citebrackets{GWgapribbonsLouie} 
using periodic boundary conditions and a plasmon-pole model. 
The effect of eigenvalue-selfconsistency, while still substantial, is  smaller for the transport gap,
leading to a ev$GW$@PBE value of 3.8\,eV.
In order to enable comparison with experiments~\citebrackets{BoundstateBoron}, where the GNRs are physisorbed on the highly polarizable Au(111) surface, 
we include the effect of the screening by the substrate 
via an image charge model devised specifically for the case of GNRs on noble
metal surfaces~\citebrackets{Meunier}.
The gap of the pristine GNR reduces by $\Delta_\text{IC}\approx 1.3\,$eV 
to $\approx 2.5$\,eV  in good agreement with
previous experimental and theoretical work~\citebrackets{RuffieuxAGNR7,wang2016giant,Meunier}.

\begin{figure}[t]
\centering
\includegraphics[width=0.47\textwidth]{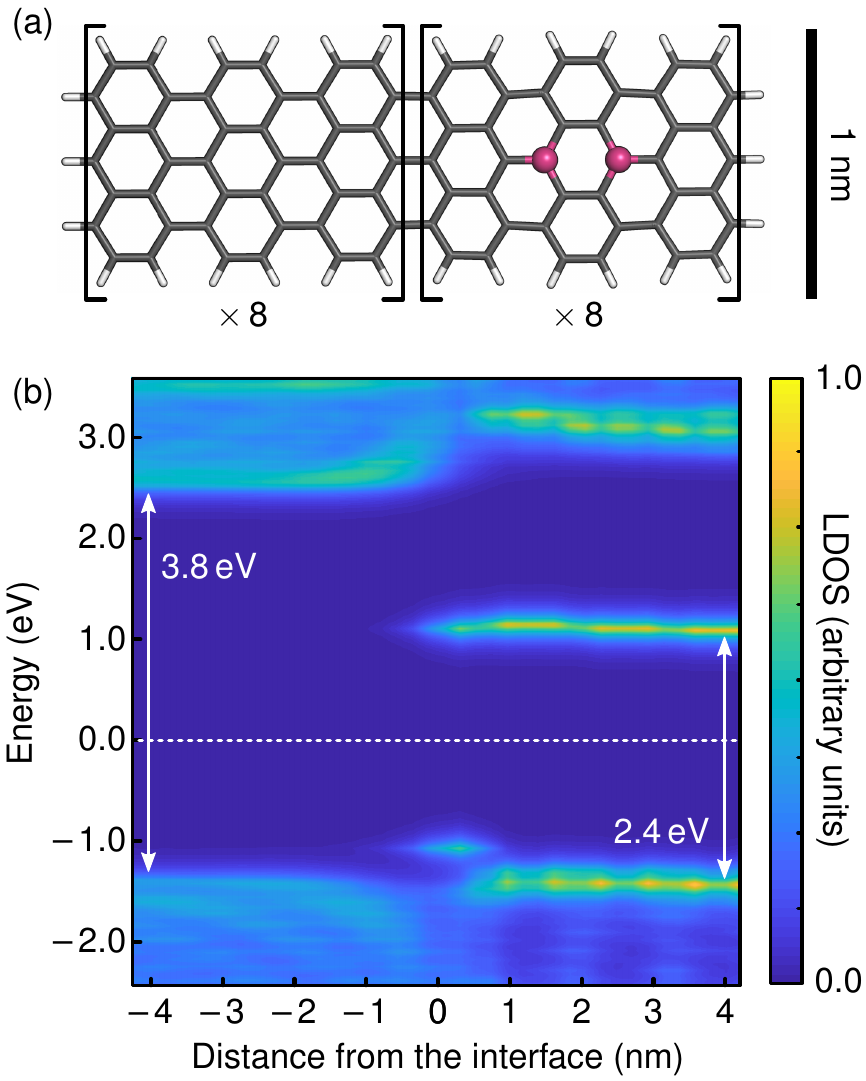}
\caption{
Graphene nanoribbon heterojunction. (a) Atomistic model including 870 atoms (boron dopants shown in purple).
(b) Local density of states across the junction
based on ev$GW$@PBE eigenvalues, with Gaussian broadening of 0.1\,eV and integrated over the plane orthogonal to the GNR axis.
The energy zero was chosen as the center of the gap.
}
\label{heteroj}
\end{figure}
Next, we turn our attention to heterostructures between doped and undoped GNRs
that have recently been demonstrated via on-surface synthesis~\citebrackets{ribbonheterojunctions,Bdopingribbon,BoundstateBoron}.
The controlled modulation of the band structure and charge carrier concentration through doping,
as well as the synthesis of atomically precise heterojunctions
are crucial milestones on the path towards graphene nanoribbon electronics.
While many-body perturbation theory in the $GW$ approximation is well-equipped to capture the level alignment, energy gaps and local density of states (LDOS) across such interfaces, 
the long range of the Coulomb interaction can make it necessary
to treat large numbers of atoms in order to obtain converged results.
Fig.~\ref{heteroj}\,(a) depicts an interface between a pristine GNR and its boron-doped variant,
as \red{realized experimentally via bottom-up synthesis} in Ref.~\citenum{Bdopingribbon}.
We perform ev$GW$@PBE calculations for the heterojunction containing 870 atoms, which converges
the gap to $\approx$\,0.1\,eV, cf.~Fig.~\ref{computed_zz_AC_gap}\,(d).
The LDOS at the interface between pristine and doped side is shown in Fig.~\ref{heteroj}\,(b).
For the bulk gap on the pristine side, we recover the value of 3.8\,eV from Fig.~\ref{computed_zz_AC_gap}\,(d), 
while \red{the empty $p$ orbitals of B give rise to a weakly dispersing acceptor 
band~\citebrackets{Bdopingribbon}, yielding a lower band gap of 2.4\,eV for the doped GNR}.
The LDOS also reveals information specific to the interface: 
the valence band maxima of the pristine and doped GNR align,
\red{making this a type-I (straddling gap) heterojunction.
Despite the perfect lattice match,}
an interface state appears close to the Fermi edge~\citebrackets{BoundstateBoron}, which can \red{introduce backscattering and thus} strongly affect the current response of the heterojunction at low bias voltages~\citebrackets{PRBPaper}.
As pointed out by Cao \emph{et al.}~\citebrackets{Cao2017},
the presence this interface state can be deduced from topological arguments, 
since the interface is one between 
a one-dimensional topological insulator ($Z_2=1$) on the left,
and a trivial insulator ($Z_2=0)$ on the right.

In summary, we have presented an efficient algorithm for computing quasiparticle energies in the $GW$ approximation,
requiring $\mathcal{O}(N^3)$ operations and $\mathcal{O}(N^2)$ memory. 
The method is a reformulation of the $GW$ space-time method~\citebrackets{AC1} in a Gaussian basis using sparse linear algebra and minimax grids~\citebrackets{liu2016cubic} for imaginary time and frequency.
Both $G_0W_0$ and eigenvalue-selfconsistent $GW$ are supported,
using either periodic or aperiodic boundary conditions.
We have implemented the algorithm in version 5.0 of the open-source CP2K package~\citebrackets{reviewJuerg}
and benchmarked its accuracy on the complete $GW$100 set of molecules, 
finding good agreement with reference implementations. 
The scalability of the algorithm was demonstrated by
computing quasiparticle energies of graphene nanoribbons containing up to 1734 atoms and the spatially resolved local density of states of a graphene nanoribbon heterojunction. 
By reducing the cost of computing accurate electron removal and addition energies in nanostructures, molecules and their composites, our work provides yet another stepping stone 
on the path towards \emph{in silico} materials design.

\section*{Computational Methods}
For the $GW$100 benchmark set~\citebrackets{GW100}, we solve the 
all-electron Kohn-Sham (KS) equations in the Gaussian and augmented plane waves scheme (GAPW)~\citebrackets{GAPW} as implemented in CP2K~\citebrackets{reviewJuerg}. 
The molecular orbitals are expanded in a def2-QZVP Gaussian-type basis~\citebrackets{GW100} [Eq.~\eqref{expGauss}],
while $GW$ quantities are expanded in  a \red{cc-pV5Z-RI} auxiliary basis~$\{\varphi_P\}$
\red{taken from the EMSL database~\citebrackets{EMSL}}.
\red{For the 17 elements from K to Ne not covered by the cc-pV5Z-RI basis, we constructed a large RI basis containing 124 sets up to I functions.}
%
12-point minimax grids were used in imaginary time and frequency.
\blue{For the analytic continuation, we construct the Pad\'{e} approximant 
on the subset of imaginary frequency points in the interval $i[0,\pm10\,\text{eV}]$,  where $+/-$ applies to virtual/occupied MOs.}

For the GNRs, we solve the singlet open-shell KS equations 
in the Gaussian and plane waves scheme (GPW)~\citebrackets{GPW} using Goedecker-Teter-Hutter pseudopotentials~\citebrackets{GTH}.
The molecular orbitals are expanded in an aug-DZVP Gaussian-type basis which converges the HOMO-LUMO gap within a few tens of meV, see also Ref.~\onlinecite{GWCP2K}.
As the auxiliary basis, we employ the corresponding RI-aug-DZVP basis from Ref.~\onlinecite{GWCP2K} which has been generated by optimizing the RI-MP2 energy to match the MP2 energy~\citebrackets{MauroRI,RIWeigend}.
For the $GW$ calculations, atom blocks of basis functions with a Frobenius norm lower than $10^{-11}$
were filtered~\citebrackets{dbcsr} to make sparse tensor operations [Eqs.~\eqref{NsquareRPA} and ~\eqref{corrSEtaurewrite}]  efficient.
This filter threshold is low enough to affect the
$G_0W_0$ HOMO-LUMO gap of the 6-anthene by less than $0.01$\,eV. 
\red{We note that using the same filter threshold for the $GW$100 set results in no filtering at all due to the small size of the molecules, i.e. the GNRs and the molecules are treated equally in this respect.} 
\blue{Further information on the choice of the filter threshold can be found in the Supporting Information. }
%
%
%
%
Again, we use 12-point minimax grids in time and frequency.
Fig.~\ref{heteroj}\,(b) was produced by projecting the LDOS onto the atomic orbitals of the GNR and summing over all nine atoms in a vertical line. 
In this way, the LDOS is integrated over the plane perpendicular to the ribbon axis.

An exemplary, annotated input file is provided in the Supporting Information on page~S6.

\section*{Acknowledgement}

We thank R.~Fasel and P.~Ruffieux for helpful discussions and M.~J.~van Setten for sharing basis sets to perform the $GW$100 benchmark.
Calculations were enabled by 
the Swiss National Supercomputing Center (CSCS), 
under 
projects ID mr2 and uzh1.
PRACE project 2016153518 is acknowledged.
This research was supported by the NCCR MARVEL, funded by the Swiss National Science Foundation.

\section*{Supporting Information}
A detailed comparison between low-scaling and canonical $\mathcal{O}(N^4)$-scaling $GW$ in a Gaussian basis including a discussion of the resolution of the identity is given; all values for the $GW$100 test set, a discussion on choosing filter parameters for sparse tensor operations, an exemplary input file together with basis sets and a discussion on the basis set convergence are reported. This
material is available free of charge via the Internet at http://
pubs.acs.org.

\bibliography{Literature}

\end{document}